\documentclass[10pt,twocolumn,english,aps,prd,superscriptaddress,tightline,nofootinbib,numerical,floatfix]{revtex4-2}
\usepackage[latin9]{inputenc}
\usepackage{babel}
\usepackage{amsmath}
\usepackage{amssymb}
\usepackage{graphicx}
\usepackage{csquotes}
\usepackage[unicode=true,
 bookmarks=false,
 breaklinks=false,pdfborder={0 0 1},backref=false,colorlinks=false]
 {hyperref}

\makeatletter

\DeclareTextSymbolDefault{\textquotedbl}{T1}

\usepackage{babel}

\usepackage{bm}

\makeatother

\begin{document}
\title{Entanglement Witness for the Weak Equivalence Principle}
\author{Sougato Bose}
\affiliation{Department of Physics and Astronomy, University College London, Gower
Street, WC1E 6BT London, UK}
\author{Anupam Mazumdar}
\affiliation{Van Swinderen Institute, University of Groningen, 9747 AG, The Netherlands}
\author{Martine Schut}
\affiliation{Van Swinderen Institute, University of Groningen, 9747 AG, The Netherlands}
\author{Marko Toro\v{s}}
\affiliation{School of Physics and Astronomy, University of Glasgow, Glasgow, G12
8QQ, UK}
\begin{abstract}
The Einstein equivalence principle is based on the equality of gravitational
mass and inertial mass, which has led to the universality of a free-fall
concept. The principle has been extremely well tested so far and has
been tested with a great precision. However, all these tests and the
corresponding arguments are based on a classical setup where the notion
of position and velocity of the mass is associated with a classical
value as opposed to the quantum \textit{entities}. Here, we will provide
a simple protocol based on creating large spatial superposition states
in a laboratory to test the \textit{fully quantum regime of the equivalence
principle} where both matter and gravity are treated at par as a quantum
entity. We will argue that such a quantum protocol is unique with
regard to testing especially the generalization of the weak equivalence
principle via witnessing quantum entanglement. 
\end{abstract}
\maketitle

\section{Introduction}

Over the last century, the general theory of relativity
has  passed a number of stringent experimental
tests~\citep{Clifford}. Among its core tenets, still experimentally
unchallenged, is the Einstein equivalence principle (EEP). The EEP,
in its {modern form} \citep{Dicke}, consists of three parts: the universality
of free fall, also known as the weak equivalence principle (WEP), local
Lorentz invariance (LLI) and local position invariance (LPI). The
WEP implies that all objects fall at the same rate, regardless of their
internal composition or structure, as long as tidal effects can be
neglected. 

Generally, the equivalence principle states as follows:
	The equations of motion for matter coupled to gravity are locally 
	identical to the equations of motion for matter in the absence of 
	gravity.

In Newton's theory of gravity (which is non-relativistic), WEP is phrased as the equality
of the inertial and gravitational mass---the mass appearing in Newton's
second law and the Newtonian gravitational potential, respectively.
In addition, LLI and LPI assume that any local non-gravitational experiment
will give the same result, regardless of the velocity of the freely-falling
reference frame in which it is performed and regardless of where
and when the experiments are performed.

The WEP has been put under experimental scrutiny from the days of
Galileo, and space-based experiments such as MICROSCOPE have placed
stringent bounds on the universality of free-fall, constraining the
value of the E\"{o}tv\"{o}s parameter to one part in $10^{15}$~\citep{MICROSCOPE:2019jix}.
Experiments with trapped atoms and ions have tested possible LLI violations,
parameterized by $\delta=|c^{-2}-1|$, with $c$ the speed of light,
confining the values of $\delta$ below $3\times10^{-22}$ \citep{atomion,Lamoreaux,Chupp}.
Atomic clock experiments, which test deviations from the gravitational
red-shift formula, $z=(1+\alpha)\frac{\Delta U}{c^{2}}$, where $z$ is
the red-shift and $\Delta U$ the Newtonian potential difference between
two clocks, have found that the LPI violation parameter $\alpha$
must be smaller than one part in $10^{6}$ \citep{peil}. 

The EEP is however formulated in an ostensibly classical framework,
and its generalization to quantum mechanics requires careful considerations~\citep{Giulini}.
Quantum systems are described in terms of wave-functions, which do not
have a point-like support and thus do not conform to the notion of
test particles, central in the formulation of EEP. As first demonstrated
by the Collela--Overhauser--Werner (COW) experiment~\citep{COW}, quantum
experiments at the interface with gravity can no longer be described
solely in terms of classical trajectories but require the computation
of the quantum phases~\citep{Aharanov,Hohensee,Overstreet}. 

In the quantum domain, where the notion of particles and trajectories becomes 
vague, matter-waves act as \emph{quantum probes} of the background 
gravitational field, requiring the generalization of the equivalence principle to the 
quantum domain~\citep{Claus,Viola,Paris,Seveso1,Seveso2,Rosi,Hu}, with ongoing experimental effort~\citep{Dimopoulos:2006nk,Roura,Asenbaum:2020era,Overstreet:2017gdp}.

However, when the gravitational field is sourced by a quantum object,
the assumption of a classical background of a gravitational field becomes
problematic, and the EEP formulations discussed above cannot be directly
applied. Some have even questioned the validity of EEP when \emph{quantum
sources} of gravity are involved~\citep{Penrose,Penrose2}, while some have pointed
out its consistency and introduced generalized notions~\citep{Hardy,Giacomini:2020ahk,Giacomini,Zych,Marletto:2020agp,Pipa:2018bui,Paunkovic:2022flx}.

To date, there is no experimental evidence about the quantum-gravity
interface in this regime, and any experiment shedding light on the
gravitational field generated by a quantum source would be a major
milestone~\citep{Bose:2017nin, Marshman:2019sne,Bose:2022uxe}.

The aims of this paper are to (i) introduce a generalized WEP
capable of testing the equivalence between the inertial and the gravitational mass with quantum sources and quantum-natured gravity
and (ii) provide a protocol for an experimental realization with matter-waves. 

 {In this paper, we put forward a quantum protocol to test the equivalence between the inertial and the gravitational mass, where the gravitational mass is sourced solely by the two quantum systems.}
We work in the framework of perturbative canonical quantum gravity
coupled to non-relativistic matter where such a problem can be unambiguously
formulated and offer a pathway for experimental implementation with
nano-particles. Using the ostensibly quantum notion of entanglement,
we introduce the notion of the \emph{entanglement entropy weak equivalence
principle} (EEWEP), which can be tested by adapting the recently proposed
quantum gravity induced entanglement of masses (QGEM) protocol~\citep{Bose:2017nin}; see also~\cite{Marletto} for a similar proposal.
Unlike previous WEP tests, which relied on classical notions or single-particle
interference, the EEWEP relies on two-particle entanglement---a hitherto
unexplored regime of the quantum--gravity interface.

%%%%%%%%%%%%%%%%%%%%%%%%%%%%%%%%%%%%%%%%%%
\section{QGEM Scheme}
\begin{figure}[h]
%\centering
\includegraphics[width=0.7\linewidth]{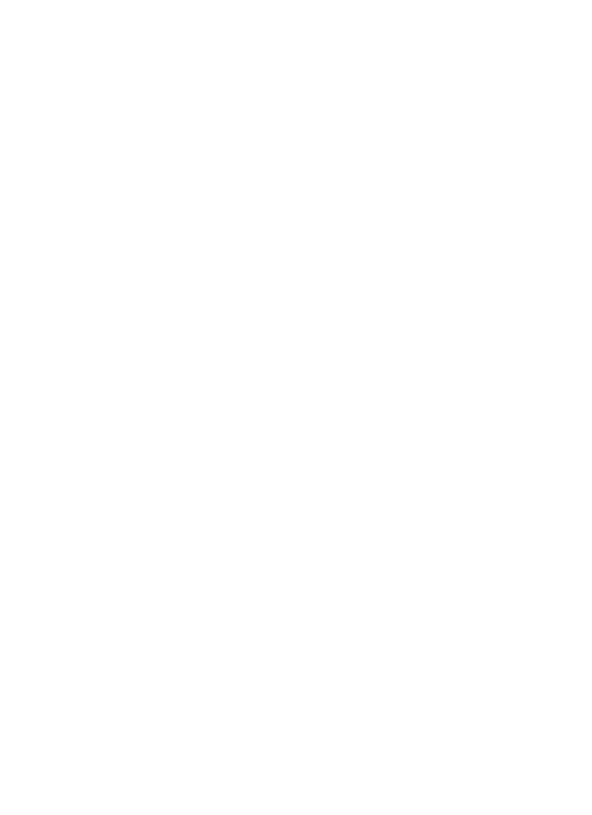} 
\caption{{Configuration where the two neutral and massive} %MDPI: We moved figure 1 behind its first citation, please confirm.
%MS: Thank you, this is fine.
 spatial superpositions
are in free-fall with the splitting $\Delta x$ and separated by
a distance $d$. The two spin states (up and down) are embedded in
the nano-crystals. The splitting between the two massive spin states
is created by an external inhomogeneous magnetic field, similar to
the Stern--Gerlach protocol. After the one-loop interference is completed,
the spin correlations are computed to witness the entanglement between
the two systems induced by the mutual quantum-natured gravitational
interaction.}
\label{configuration}
\end{figure}

Quantum mechanical sources of gravity pose significant
conceptual questions and have led to several approaches to quantum
gravity~\citep{Kiefer}. Here, we work within the framework of
the perturbative low energy effective field theory of quantum gravity coupled to the non-relativistic
matter-waves~\citep{Gupta-1952,Gupta,Donoghue:1994dn}. Within this framework, the
QGEM protocol aims to test the \textit{quantum nature for gravity}
in a laboratory with the basic blueprint shown in Figure~\ref{configuration}~\citep{Bose:2017nin,Marletto}.
In a nutshell, the two particles are placed sufficiently far apart that the
electromagnetic interactions are negligible and at the same time
close enough that the two particles become entangled through the gravitational
interaction. The underlying mechanism for the generation of entanglement
has been analysed within low energy effective field theory~\citep{Marshman:2019sne,Bose:2022uxe} and
the framework of the Arnowitt--Desse--Meissner (ADM) approach~\citep{Danielson},
as well as in the path integral approach~\citep{Christodoulou2}.
In the language of effective field theory of quantum gravity, at low energies, the two quantum systems are entangled due to the exchange of a graviton
containing both the spin-2 and the spin-0 components, which are the
dynamical off-shell degrees of freedom of a massless graviton in four dimensions~\citep{Bose:2022uxe,Marshman:2019sne}.

The QGEM protocol, assuming the locality of interactions, results
in an entanglement witness for the quantum character of gravity.
The result is concurrent to the Local Operation and Classical Communication
(LOCC) theorem \citep{LOCC}. The LOCC states that the two quantum
systems cannot be entangled via a classical channel if they were not
entangled to begin with, or entanglement cannot be increased by local
operations and classical communication. Therefore, by witnessing the
entanglement between the two masses, specifically by detecting quantum
correlations between the spins that are embedded in the two test
masses, we can ascertain whether the gravity is a classical or a quantum
entity.

In the QGEM framework, the two free-falling particles thus interact
gravitationally in an ostensibly quantum regime---each particle
is placed in a superposition and acts as a \emph{quantum source} for
the gravitational field. The free-fall of
the left (right) particle is determined by the non-classical gravitational
field generated by the right (left) particle. We are thus confronted
with a free-fall situation that goes beyond classical or quantum
EEP in a fixed background gravitational field and therefore requires a novel way of testing the equivalence principle, 
which we call the entanglement entropy weak equivalence principle (EEWEP). 

We define the EEWEP using the relative entanglement entropy generated
between the two gravitationally coupled particles (see Equation~(\ref{rel-ent})). We now first discuss the entanglement entropy in the QGEM scheme.

%%%%%%%%%%%%%%%%%%%%%%%%%%%%%%%%%%%%%%%%%%
\section{Entanglement Entropy}
We consider two masses with embedded spins as shown in Figure~\ref{configuration},
each of which is placed in a spatial superposition of size $\Delta x$.
The joint quantum state of the spins $\left|\Psi(0)\right\rangle =\frac{1}{2}\left(\left|\uparrow,\uparrow\right\rangle +\left|\downarrow,\downarrow\right\rangle +\left|\uparrow,\downarrow\right\rangle +\left|\downarrow,\uparrow\right\rangle \right)$
will evolve to
\vspace{-6pt}
\begin{alignat}{1}
\left|\Psi(t)\right\rangle =\frac{1}{2}e^{i\phi} & \left(\left|\uparrow,\uparrow\right\rangle +e^{i\Delta\phi_{\text{ent}}}\left|\uparrow,\downarrow\right\rangle +\left|\downarrow,\downarrow\right\rangle +e^{i\Delta\phi_{\text{ent}}}\left|\downarrow,\uparrow\right\rangle \right),
\end{alignat}
where $\phi$ is a global phase, and $\Delta\phi_{\text{ent}}$ is
the entanglement phase. It has been shown that the leading order contribution
to the entanglement phase is given by~\citep{Bose:2022uxe}
\begin{equation}
\Delta\phi_{\text{ent}}\sim\frac{2Gm_{\text{g}}^{2}t\Delta x^{2}}{\hbar d^{3}}+\cdots\,,\label{eq:dphi}
\end{equation}
where $t$ is the evolution time, $d$ is the distance between two
test masses, $G$ and $\hbar$ are Newton's and Planck's constants,
respectively, and $\cdots$ contain higher-order corrections~{{(the operator valued gravitational Hamiltonian,} %MDPI: We removed the foot note as regular text as it was not allowed in our journal, please confirm. The same as the following foot note .
% MS: Yes, confirmed.
which induces the entanglement
phase, can be computed using quantum perturbation theory. At the second
post-Newtonian order, the Hamiltonian contains momentum-dependent terms
that we assume to be negligible. Similarly, we neglect other higher-order 
curvature effects $\sim\mathcal{O}(\Delta x^{4})$ by assuming
the superposition size is small compared to the distance between the
two particles. Such assumptions are reminiscent of the assumption
of a localized point particle in the formulation of the classical
WEP, which isolates the leading order effect contributing to the acceleration
of point particles. In a complete analogy, we are considering the leading
order tidal effect shown in Equation~(\ref{eq:dphi}) that 
dominates the generation of the entanglement}). The mass $m_{g}$ that appears in Equation~\eqref{eq:dphi} is to be
identified with the gravitational mass (i.e., the mass appearing in
the coupling to gravity).

The spatial superposition is created by the electromagnetic interaction,
namely via the Stern--Gerlach protocol, which involves inhomogeneous
magnetic fields, by displacing the particle according to the spin
state. For the purpose of illustration, we assume that the size
of the superpositions is given by
\begin{equation}
\Delta x\sim\frac{f}{m_{\text{i}}}\tau_{a}^{2},\label{eq:dx}
\end{equation}
where $f$ is the force used to prepare/recombine the superpositions
in a time $\tau_{a}$. The mass $m_{i}$ appearing in Equation~\eqref{eq:dx}
is to be identified with the inertial mass (i.e., the mass in the
free-particle Hamiltonian).
For the sake of simplicity, we are assuming a simple mechanism for
creating $\Delta x$. In reality, we have to specify all the details
of how the creation and recombination of the trajectories work
out. See for details~\citep{Marshman:2021spm,Zhou:2022frl}. However, these do
not affect the inertial mass, and also all the rest of the model parameters drop out in our definition of $\eta_{s}$, to be defined in Section \ref{sec:eewep}.

Combing Equations~(\ref{eq:dphi}) and (\ref{eq:dx}), we then find 
\begin{equation}
\Delta\phi_{\text{ent}}\sim\frac{2Gtf^{2}\tau_{a}^{4}}{\hbar d^{3}}\left(\frac{m_{\text{g}}}{m_{\text{i}}}\right)^{2}+\cdots,\label{eq:dphi2}
\end{equation}
where $\cdots$ again contain higher-order corrections. From Equation~(\ref{eq:dphi2}),
we see that the entanglement phase depends on the fundamental constants
($G$ and $\hbar$), on parameters that can be controlled by the experimentalists
($t$, $\tau_{a}$, $d$ and $f$) and finally on the ratio $m_{\text{g}}/m_{\text{i}}$.%MS: added a dot to end the sentence here.

{Of course, the two masses in systems} %MDPI: . MS: I made a separtely paragraph out of what used to be a footnote, to clean up the text. And I removed the paratheses.
 1 and 2 need not be the same,
allowing us to modify the above expression to 
\begin{equation}
\Delta\phi_{\text{ent}}\sim\frac{2Gtf^{2}\tau_{a}^{4}}{\hbar d^{3}}\left(\frac{m_{\text{g}}^{(1)}}{m_{\text{i}}^{(1)}}\frac{m_{\text{g}}^{(2)}}{m_{i}^{(2)}}\right)\frac{(m_{\text{i}}^{(1)}+m_{i}^{(2)})^{2}}{m_{\text{i}}^{(1)}m_{i}^{(2)}}\, ,
\label{phio}
\end{equation}
where $m_{\text{g}}^{(j)}$ and $m_{\text{i}}^{(j)}$ denote the gravitational
and inertial mass of the $j$-particle, respectively. For the purpose
of illustration, we take the inertial masses for both the systems
to be the same, but the final expression for the relative entanglement
entropy in Equation~\eqref{eta2} remains the same also in the case of
unequal masses.

{It is instructive to compare the phase in} %MDPI: . MS: I put what used to be a footnote in a new paragraph to make the text cleaner, and removed the parentheses.
 Equation~\eqref{eq:dphi2} with the phase obtained in the COW experiment. Specifically, the  COW phase is given by $\phi_\text{COW}= m_g g \Delta x t/\hbar$, where $m_g$ is the gravitational mass, $g$ is the Earth's gravitational acceleration, and $\Delta x$ ($t$) is the superposition size (evolution time) \cite{Sakurai,Rauch}. We now use again \eqref{eq:dx} to obtain
\begin{equation}
    \phi_{\text{COW}}=  \frac{G M f t \tau_a^2}{\hbar R^2} \left(\frac{m_\text{g}}{m_\text{i}}\right),
\end{equation}
 where we have inserted $g=G M/R^2$ ($M$ is the gravitational mass of the Earth, and $R$ is the distance between the experiment and the center of the Earth). The COW phase can thus be used to discern between the gravitational and inertial mass, albeit with a different scaling of the ratio $m_\text{g}/m_\text{i}$. More importantly, the COW experiment is conceptually different to the situation depicted in Figure~\ref{configuration}. The COW phase arises from the classical background gravitational field generated by the Earth, with the experimental setup bound to its surface. Rather, in the case of Equation~\eqref{eq:dphi2}, the whole experiment is in free-fall (and hence the COW phase is absent), and the gravitational field is sourced by the particles themselves, each of which is prepared in a superposition state.

Using a similar parametrization as in the classical WEP tests, we quantify
the deviation from the expected behavior %MS: changed spelling of behaviour to be consistent with the rest of the paper
by writing 
\begin{equation}
\frac{m_{\text{g}}}{m_{\text{i}}}\equiv1+\xi,\label{eq:xi}
\end{equation}
where $\xi$ is a dimensionless parameter that can be extracted experimentally.
If we have\linebreak $m_{\text{g}}=m_{\text{i}}$, then the entanglement phase
will not depend on the ratio between the gravitational and inertial
mass, i.e., $\Delta\phi_{\text{ent}}=2Gtf^{2}\tau_{a}^{4}/(\hbar d^{3})$. 

To quantify how the degree of entanglement changes with the parameter
$\xi$, we combine Equations~\eqref{eq:dphi2} and \eqref{eq:xi} and compute
the entanglement entropy $S_{\xi}(t)$. Assuming that the entanglement
phase $\Delta\phi_{\text{ent}}$ is small, we find a simple expression
\begin{equation}
S_{\xi}(t)=\frac{f^{4}G^{2}\tau_{a}^{8}t^{2}}{2d^{6}\hbar^{2}}(1+\xi)^{4}.\label{eq:sxi}
\end{equation}
By comparing Equations~\eqref{eq:dphi2} and \eqref{eq:sxi}, we see that
the entanglement entropy is simply the square of the entanglement
phase, i.e., $S_{\xi}(t)\sim\Delta\phi_{\text{ent}}^{2}/8$. It is interesting to note that the entanglement entropy is proportional to the $G^2/\hbar^2$ contribution.

% %%%%%%%%%%%%%%%%%%%%%%%%%%%%%%%%%%%%%%%%%%
\section{EEWEP}\label{sec:eewep}
We can define the \emph{relative} entanglement entropy, similar to
the E\"{o}tv\"{o}s  parameter, 
\begin{equation}
\eta_{s}(t)\equiv\left|\frac{S-S_{\xi}(t)}{S+S_{\xi}(t)}\right|,\label{rel-ent}
\end{equation}
where $S\equiv S_{0}(t_{\text{\text{ref}}})$ corresponds to the entanglement
entropy with $\xi=0$ computed at a reference time $t_{\text{\text{ref}}}$.
Combining Equations~\eqref{eq:sxi} and \eqref{rel-ent}, we then find
\begin{equation}
\eta_{s}(t)=\left|\frac{t_{\text{ref}}^{2}-(1+\xi)^{4}t^{2}}{t_{\text{ref}}^{2}+(1+\xi)^{4}t^{2}}\right|,\label{eta2}
\end{equation}
where all the fundamental constants and the experimental parameters on
the right-hand side have cancelled, apart from two times, $t$
and $t_{\text{ref}}$, and the parameter $\xi$. 

\begin{figure}[h]
\includegraphics[width=\linewidth]{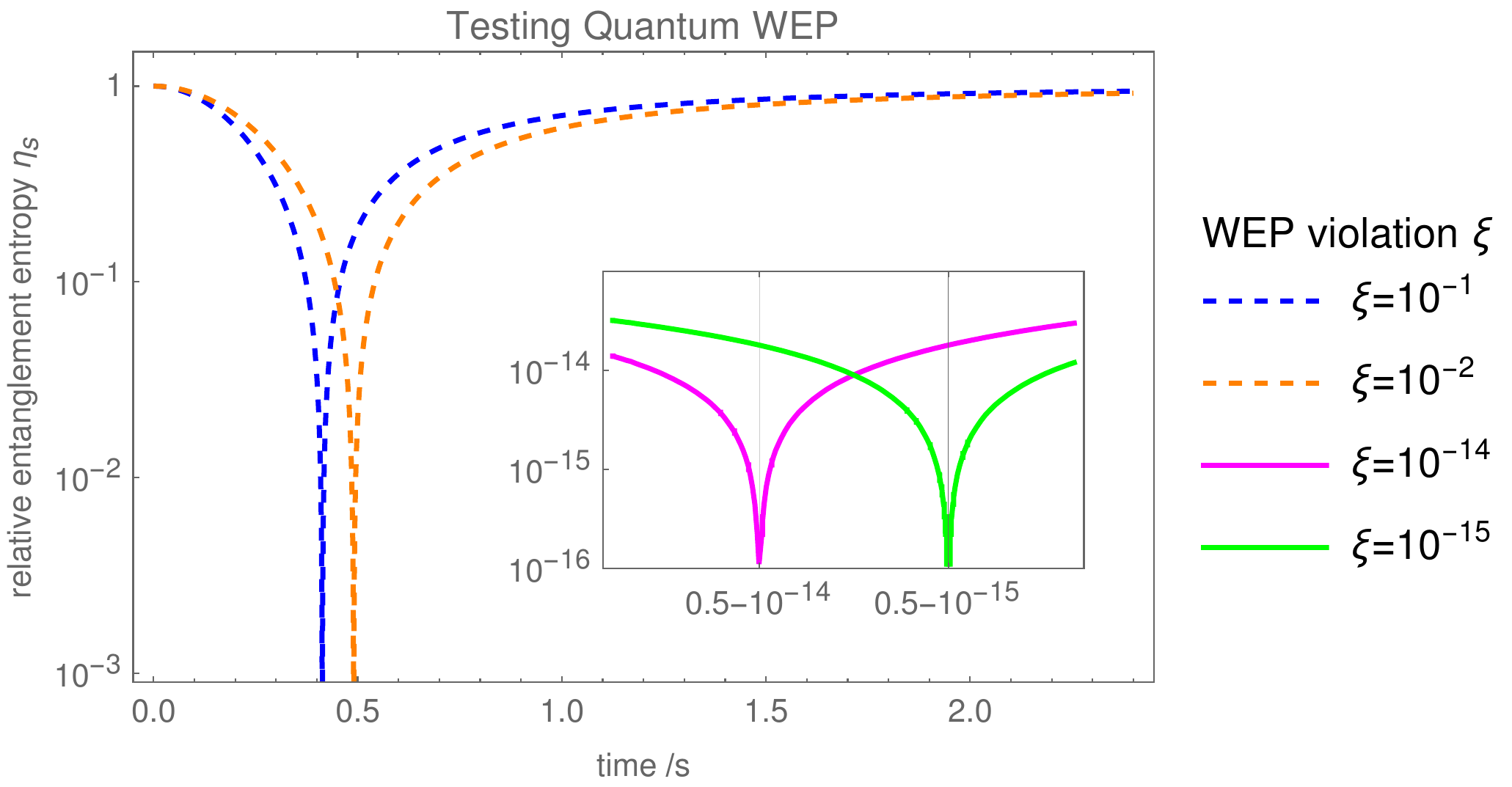} \caption{Plot of relative entanglement entropy $\eta_{s}(t)=\vert(S-S_{\xi}(t))/(S+S_{\xi}(t))\vert$
with respect to the time evolution $t$. $S_{\xi}(t)$ is the entanglement
entropy with the EEWEP violation given by $\xi=m_{\text{g}}/m_{\text{i}}-1$,
and $S=S_{0}(t_{\text{ref}})$ is the entanglement entropy without
the EEWEP violation computed at the reference time $t_{\text{ref}}=0.5~\text{s}$.
Different colours exhibit different values of the EEWEP violation $\xi$.
Testing the EEWEP down to $\xi\sim10^{-2}$ is within experimental
possibilities and can be accomplished by measuring the relative entanglement
entropy $\eta_{s}$ with accuracy $10^{-2}$. Testing the EEWEP to
one part in $10^{2}$ would probe a hitherto unexplored quantum notion
of free-fall distinct from any classical test of WEP.  The
inner embedded plot shows again the relative entropy with respect
to time. As an illustration of the scaling of the experimental requirements,
we consider the more ambitious value $\xi\sim$10$^{-15}$. We would
require $\sim\text{fs}$ resolutions, achievable with atomic clocks,
and a scheme for determining the relative entanglement entropy with
accuracy $10^{-15}$, the latter being beyond current experimental
possibilities.}
\label{fig:entropy-1}
\end{figure}

We are thus led to define the entanglement entropy weak equivalence
principle (EEWEP): \emph{the relative entanglement entropy generated
between two gravitationally coupled particles is independent of their
structure or composition}.
% added: ---------------------------------------------------------------------------------------------------------------------------------------
{The EEWEP is related to the Newtonian formulation} %MDPI: . MS: Separated the foootnote from the previous sentence and removed the parentheses.
 of the WEP, because the 
measure $\eta_s$ of the EEWEP will provide a value of $\xi$, which is a measure for the 
equivalence of inertial and gravitational mass as was the WEP in the Newtonian domain.
The EEWEP however provides a reformulation of the Newtonian WEP principle that does 
not use statements about trajectories, which are debatable in the quantum domain, but 
provides a purely quantum reformulation of the WEP.
%until here  ---------------------------------------------------------------------------------------------------------------------------------------
Now, $\eta_{s}$ is the measure of EEWEP, and $\xi$ denotes the violation of EEWEP. 
To measure precisely the EEWEP violation
$\xi$, one requires only measurements of the entanglement entropy
$S_{\xi}$ at the times $t$ and $t_{\text{ref}}$---one first obtains
the value of $\eta_{s}$ from Equation~\eqref{rel-ent}, and then combining
it with Equation~\eqref{eta2} extracts the value of the WEP violation
$\xi$. Figure~\ref{fig:entropy-1} shows the time-dependent evolution
of the relative entanglement entropy for different values of $\xi$.

In the first place,  probing the EEWEP violation up to  $\xi\sim$10$^{-2}$,
is well within the experimental possibilities, as it only requires temporal control with accuracy $\sim$10 ms 
and measurement of entanglement entropy to $S_{\xi}\sim{\cal O}(10^{-2})$. 
For the purpose of  illustration, we have also considered experimental values such as $\xi\sim10^{-15}$ just to have an indicative comparison with
the sensitivities achieved in the classical WEP tests{~\citep{MICROSCOPE:2019jix}}. Measuring the entanglement
entropy with the accuracy $S_\xi\sim {\cal O}(10^{-15})$ would require exquisite control
of the experiment and a large number of experimental runs, which is
currently beyond experimental realities. Nonetheless, recent advancements
of keeping track of the frequency ratio measurements up to $18$-digit accuracy
may be the way to track the time evolution of the relative entanglement
entropy~\citep{BACON}. Further, note that the time intervals can
be controlled with a great precision given the historical achievement
of pico-second ($10^{-12}$ s) pulse rise/fall timings with microwave
lasers \citep{pulse1}, with femto-second timings also achieved more recently---see \citep{pulse2}---which can, in principle, be used to control an
interferometer.
{Quantum tests of the classical WEP (performed by comparing the acceleration between different atoms/isotopes in the earth's gravitational field) have reached an experimental accuracy of $\xi\sim10^{-7}$~\citep{Tarallo:2014oaa,Schlippert:2014xla,Fray:2004zs}. 
Matching this accuracy would require measurements of entanglement entropy with an accuracy of $S_\xi \sim {\cal O}(10^{-7})$, which have not been realized experimentally.
}
%Although this would require a very good control, entanglement entropy has been measured with an accuracy $1$ part in $10^{5}$ in 2018~\citep{}.

In principle, one could be able to test the EEWEP for any
massive superpositions, but in practice, %MS: adjusted spelling to be consistent with the rest of the text.
one is limited to experiments where the interaction between 
the two particles is dominated by gravity. To
witness gravitational-induced entanglement, we need the gravitational
interaction to be dominant over other known Standard Model interactions,
such as the electromagnetic-induced interactions. This requires a massive
superposition of order $m_{A}\sim m_{B}\sim{\cal O}(10^{-15}-10^{-14})$kg,
$\Delta x\sim10-100~{ \mu \text{m}}$, and d $\sim400~{\mu \text{m}}$, as
suggested in the original QGEM paper~\citep{Bose:2017nin} and in
\citep{vandeKamp:2020rqh}. These parameters will generate an appreciable
graviton-induced entanglement phase and will dominate over the photon
induced Casimir--Polder potential for neutral masses. The mass range
of the quantum system is such that it effectively modifies the graviton
vacuum by less than one-graviton excitation~\citep{Bose:2021ekc,Bose2022},
and the emission of any gravitational waves is indeed negligible~\citep{toros,Sabbata}.

There are however still many experimental challenges: cooling the
nano-crystal~\citep{Delic,Magrini1,Tebbenjohanns1}, creating superpositions~\citep{Scala,Wan1,Folman,Zhou1,Margalit1,Margalit:2020qcy,Marshman:2021spm,Pedrnales},
tackling decoherence {(both from ``standard'' sources of decoherence such as collisions with air molecules~\citep{romero,Bose:2017nin,vandeKamp:2020rqh,Tilly:2021qef,Schut1} as well as from the gravitational coupling to classical or quantum detectors~\citep{Gunnink:2022ner,Torrieri:2022znj})},
as well as controlling noise sources~\citep{Toros:2020dbf}.

% %%%%%%%%%%%%%%%%%%%%%%%%%%%%%%%%%%%%%%%%%%
\section{Discussion}
To summarize% MS: changed the spelling of summarise to summarize to be consistent with the rest of the paper
, we have provided the first
ever quantum protocol to probe the concept of a free-fall in the framework of
perturbative quantum gravity coupled to the non-relativistic quantum matter.
We have used the graviton-induced entanglement between the two particles
to define the concept of EEWEP---a bonafide quantum test of a free-fall with quantum sources
of gravity. To quantify EEWEP violations, we have introduced the parameter
$\xi$, {which measures the difference between the gravitational mass and the inertial mass}, and pointed out that the violation of EEWEP within $\xi \sim {\cal O}(10^{-2})$ can perhaps be tested in a near-future experiment.

There are also future avenues to probe the concept of free-fall with
quantum sources of gravity. Typically, in any theory where the constants
of nature are replaced by dynamical entities, they will tend to violate
the equivalence principle~\citep{Damour:2012rc}. Therefore, the Brans--Dicke
theory of gravity~\citep{Brans} and string theory would violate
the weak equivalence principle~\citep{Mende:1992pm,Damour:2002mi}.\linebreak
This is due to the fact that Newton's constant depends on the
running dilaton, which means that in the gravitational sector, there
will be new dynamical off-shell degrees of freedom. This will also be the
case in the context of higher derivative theories of gravity~\citep{Van,Biswas2}
and non-local theories of gravity~\citep{Biswas:2011ar,Biswas3,Tomboulis,Modesto1,Edholm:2016hbt}.
It will be interesting to study in the %MS: added 'the'
future the predictions for the EEWEP violations in such theories.

%%%%%%%%%%%%%%%%%%%%%%%%%%%%%%%%%%%%%%%%%%%%%%

\section*{Acknowledgements:} SB would like to acknowledge EPSRC grants
No. EP/N031105/1 and EP/S000267/1. AM's research is funded by the
Netherlands Organisation for Science and Research (NWO) grant number
680-91-119. M.S. is supported by the Fundamentals of the Universe
research programme within the University of Groningen. MT acknowledges
funding by the Leverhulme Trust (RPG-2020-197).

%%%%%%%%%%%%%%%%%%%%%%%%%%%%%%

\end{document}